\documentclass[prl,aps,twocolumn,floatfix,showpacs]{revtex4}
\usepackage{graphics}
\begin{document}
\title{Trapped fermion mixtures with unequal masses: a Bogoliubov-de Gennes approach}
\author{M. Iskin and C. J. Williams}
\affiliation{Joint Quantum Institute, National Institute of Standards and Technology, and 
University of Maryland, Gaithersburg, Maryland 20899-8423, USA.}
\date{\today}

\begin{abstract}

We use the Bogoliubov-de Gennes formalism to analyze the ground state phases 
of harmonically trapped two-species fermion mixtures with unequal masses. 
In the weakly attracting limit and around unitarity, we find that the superfluid 
order parameter is spatially modulated around the trap center, and that its global 
maximum occurs at a finite distance away from the trap center where the 
mixture is locally unpolarized. As the attraction strength increases 
towards the molecular limit, the spatial modulations gradually disappear while 
the Bardeen-Cooper-Schrieffer (BCS) type nonmodulated superfluid region 
expands until the entire mixture becomes locally unpolarized.

\end{abstract}

\maketitle

The many-body physics of fermion mixtures with mismatched Fermi surfaces
has been a longstanding problem for many researchers ranging from the condensed 
and nuclear matter to the high energy and astrophysics communitites~\cite{casalbuoni}.
While there are various theoretical proposals for the ground state of such 
systems including the Fulde-Ferrell-Larkin-Ovchinnikov (FFLO), Sarma and 
breached pair superfluid phases, strong experimental evidence
for their observation is still lacking. Following the recent experiments 
on two-component $^6$Li mixtures with unequal populations~\cite{mit, rice}, 
a new wave of theoretical interest in this problem has been sparked in the
atomic and molecular physics communities. In these experiments the phase 
diagram of trapped (finite) systems has been studied as a function of 
population difference, temperature and two-body scattering length, showing 
superfluid and normal phases and a phase separation between them~\cite{mit-pd}.

Motivated by these experiments, phase diagrams of harmonically 
trapped mixtures with unequal populations have been extensively analyzed 
in both three- and one-dimensional systems. At the mean-field 
level in three dimensions, while fully quantum mechanical Bogoliubov-de 
Gennes (BdG) calculations provide some evidence for the FFLO type spatially 
modulated superfluid phase, such a phase is completely absent in 
calculations based on the semi-classical local density approximation 
(LDA)~\cite{mizushima, torma}. 
Therefore, it is still an open question whether these spatial modulations
are related to the FFLO superfluidity or are simply finite size effects.
However, in exactly tractable one dimensional systems, FFLO structure of the 
superfluid phase have been identified in trapped as well as infinite 
systems~\cite{1Dorso, 1Dhui, feigun, 1Dtezuka, 1Dbatrouni}. 
These works arguably suggest that the ground state of polarized mixtures is 
also an FFLO type superfluid in three dimensions 
along with the earlier BdG results~\cite{mizushima, torma}.

\begin{figure} [htb]
\centerline{\scalebox{0.55}{\includegraphics{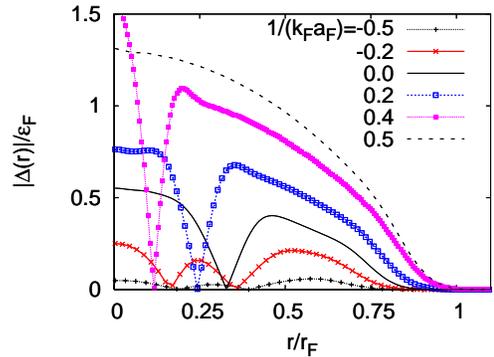} }}
\caption{\label{fig:op}
(Color online) The local superfluid order parameter $|\Delta(r)|$ versus distance 
$r$ is shown for a population balanced mixture of $^6$Li and $^{40}$K atoms. 
The BCS type superfluidity first occurs at a finite distance away from the trap 
center where the mixture is locally unpolarized. As the attraction strength increases 
towards the molecular limit, it gradually expands until the entire mixture 
becomes locally unpolarized. 
}
\end{figure}

Two-species fermion mixtures with unequal masses offer 
a very natural way of creating superfluidity with mismatched Fermi surfaces, 
and there have been increasing theoretical~\cite{liu, caldas, iskin-mixture, 
pao-mixture, lin-mixture, parish-mixture, green, orso-mixture, baranov, blume} 
and experimental~\cite{taglieber, ville} interest in studying such systems. 
For instance, $^6$Li-$^{40}$K mixtures have recently been trapped 
and interspecies Feshbach resonances have been 
identified~\cite{taglieber, ville}, opening a new frontier in ultracold atom 
research to study exotic many-body phenomena. In this manuscript, we go 
beyond the LDA method~\cite{pao-mixture, lin-mixture}, and use the 
BdG formalism to analyze the ground state phases of harmonically trapped 
$^6$Li-$^{40}$K mixtures. Our main results are as follows. 
In the weakly attracting limit and around unitarity, we find that the superfluid 
order parameter is spatially modulated around the trap center, and that its 
global maximum occurs at a finite distance away from the trap center 
where the mixture is locally unpolarized. 
As the attraction strength increases towards the molecular limit, the spatial 
modulations gradually disappear while the Bardeen-Cooper-Schrieffer (BCS) 
type nonmodulated superfluid region expands until the entire mixture 
becomes locally unpolarized as shown in Fig.~\ref{fig:op}.

We obtain these results by using the following Hamiltonian (in units of $\hbar = k_B = 1$)
\begin{eqnarray}
\label{eqn:hamiltonian}
H = \int d\mathbf{r} \left[\sum_\sigma \psi_\sigma^\dagger(\mathbf{r}) 
{\cal K}_\sigma(\mathbf{r}) \psi_\sigma(\mathbf{r}) - g \Psi^\dagger(\mathbf{r}) \Psi(\mathbf{r}) \right],
\end{eqnarray}
to describe two-component fermion mixtures with zero-ranged attractive $(g > 0)$ 
interactions, where $\psi_{\sigma}^\dagger (\mathbf{r})$ and $\psi_{\sigma} (\mathbf{r})$ 
field operators create and annihilate a pseudo-spin $\sigma$ fermion at position 
$\mathbf{r}$. Here, we have introduced the operators
$
{\cal K}_\sigma(\mathbf{r}) = -\nabla^2/(2M_\sigma) - \mu_\sigma(r)
$
and
$
\Psi(\mathbf{r}) = \psi_\downarrow(\mathbf{r}) \psi_\uparrow(\mathbf{r}),
$
where $M_\sigma$ is the mass, $\mu_\sigma(r) = \mu_\sigma - V_\sigma(r)$ is the 
local chemical potential, $\mu_\sigma$ is the global chemical potential and 
$V_\sigma(r) = M_\sigma \omega_\sigma^2 r^2/2$ is the trapping potential of 
the $\sigma$ fermions. In the mean-field approximation for the superfluid phase, 
this Hamiltonian reduces to
$
H_{MF} = \int d\mathbf{r} [
\sum_\sigma \psi_\sigma^\dagger(\mathbf{r}) {\cal K}_\sigma(\mathbf{r}) \psi_\sigma(\mathbf{r}) 
- \Delta(\mathbf{r})\psi_\uparrow^\dagger(\mathbf{r}) \psi_\downarrow^\dagger(\mathbf{r}) 
- \Delta^*(\mathbf{r})\psi_\downarrow(\mathbf{r}) \psi_\uparrow(\mathbf{r}) 
+ |\Delta(\mathbf{r})|^2/g],
$
where the self-consistency field
$
\Delta(\mathbf{r}) = g \langle \psi_\uparrow(\mathbf{r}) \psi_\downarrow(\mathbf{r}) \rangle
$
is the local superfluid order parameter and $\langle ... \rangle$ is a thermal average.

$H_{MF}$ can be diagonalized via the Bogoliubov-Valatin transformation
$
\psi_\sigma(\mathbf{r}) = \sum_\eta[u_{\eta,\sigma}(\mathbf{r}) \gamma_{\eta,\sigma} 
- s_\sigma v_{\eta,\sigma}^*(\mathbf{r}) \gamma_{\eta,-\sigma}^\dagger],
$
where $u_{\eta,\sigma}(\mathbf{r})$ and $v_{\eta,\sigma}(\mathbf{r})$ are the wavefunctions
and $\gamma_{\eta,\sigma}^\dagger$ and $\gamma_{\eta,\sigma}$ are the operators corresponding 
to the creation and annihilation of a pseudo-spin $\sigma$ quasiparticle,
and $s_\uparrow = +1$ and $s_\downarrow = -1$. This leads to the BdG equations
\begin{equation}
\label{eqn:bdg}
\left[ \begin{array}{cc}
{\cal K}_\uparrow(\mathbf{r}) & \Delta(\mathbf{r}) \\
\Delta^*(\mathbf{r}) & -{\cal K}_\downarrow^*(\mathbf{r}) 
\end{array} \right]
\mathbf{\varphi_{\eta,\sigma}}(\mathbf{r}) 
= s_\sigma \epsilon_{\eta,\sigma} \mathbf{\varphi_{\eta,\sigma}} (\mathbf{r}),
\end{equation}
where $\epsilon_{\eta,\sigma} > 0$ are the eigenvalues and $\mathbf{\varphi_{\eta,\sigma}} (\mathbf{r})$ 
are the eigenfunctions given by
$
\mathbf{\varphi_{\eta,\uparrow}}^\dagger (\mathbf{r}) 
= [u_{\eta,\uparrow}^*(\mathbf{r}), v_{\eta,\downarrow}^*(\mathbf{r})]
$
and
$
\mathbf{\varphi_{\eta,\downarrow}}^\dagger (\mathbf{r}) 
= [v_{\eta,\uparrow}(\mathbf{r}), -u_{\eta,\downarrow}(\mathbf{r})].
$
Since the BdG equations are invariant under the 
transformation $v_{\eta,\uparrow}(\mathbf{r}) \to u_{\eta,\uparrow}^*(\mathbf{r})$,
$u_{\eta,\downarrow}(\mathbf{r}) \to -v_{\eta,\downarrow}^*(\mathbf{r})$ and 
$\epsilon_{\eta,\downarrow} \to -\epsilon_{\eta,\uparrow}$, it is sufficient to solve only 
for $u_\eta(\mathbf{r}) \equiv u_{\eta,\uparrow}(\mathbf{r})$, 
$v_\eta(\mathbf{r}) \equiv v_{\eta,\downarrow}(\mathbf{r})$ and 
$\epsilon_\eta \equiv \epsilon_{\eta,\uparrow}$ as long as we keep all of the 
solutions with positive and negative eigenvalues.

We assume
$
\Delta(\mathbf{r}) = - g \sum_{\eta} u_{\eta}(\mathbf{r}) v_{\eta}^*(\mathbf{r}) f(\epsilon_{\eta})
$
is real which is sufficient to describe both nonmodulated and spatially modulated
superfluid phases. Here, $f(x) = 1/[\exp(x/T) + 1]$ is the Fermi function and 
$T$ is the temperature. This equation has an ultraviolet divergence due to 
zero-ranged interactions, and it can be regularized by relating $g$ to the 
two-body scattering length $a_F$ of the $\uparrow$ and $\downarrow$ 
fermions~\cite{regularization}. This leads to
$
1/g = -M_r/(4\pi a_F) + M_r k_c(r)/(2\pi^2) - [M_r k_F(r)/(4\pi^2)] \log\{[k_c(r)+k_F(r)]/[k_c(r)-k_F(r)] \},
$
where $M_r = 2M_\uparrow M_\downarrow/(M_\uparrow + M_\downarrow)$ is
twice the reduced mass of the $\uparrow$ and $\downarrow$ fermions,
$
k_F^2(r)/(2M_r) = [\mu_\uparrow(r) + \mu_\downarrow(r)]/2
$
and
$
k_c^2(r)/(2M_r) = \epsilon_c + [\mu_\uparrow(r) + \mu_\downarrow(r)]/2.
$
Here, $\epsilon_c$ is the energy cutoff to be specified below, and our results depend 
weakly on the particular value of $\epsilon_c$ provided that it is chosen 
sufficiently high. Furthermore, the order parameter equation has to be solved 
self-consistently with the number equations
$
N_\sigma = \int d\mathbf{r} n_\sigma(\mathbf{r}) 
$
where
$
n_\sigma(\mathbf{r}) = \langle \psi_\sigma^\dagger(\mathbf{r}) \psi_\sigma(\mathbf{r}) \rangle
$
is the local density of $\sigma$ fermions. This leads to 
$
n_\uparrow(\mathbf{r}) = \sum_{\eta} |u_\eta(\mathbf{r})|^2 f(\epsilon_\eta)
$
and
$
n_\downarrow(\mathbf{r}) = \sum_{\eta} |v_\eta(\mathbf{r})|^2 f(-\epsilon_\eta).
$

It is very natural to expand $u_\eta(\mathbf{r})$ and $v_\eta(\mathbf{r})$ in 
the complete basis of the harmonic trapping potential eigenfunctions,
which are given by the Schr\"odinger's equation
$
{\cal K}_\sigma(\mathbf{r}) \phi_{n,\ell,m}^\sigma (\mathbf{r}) 
= \xi_{n,\ell}^\sigma \phi_{n,\ell,m}^\sigma (\mathbf{r}),
$
where 
$
\xi_{n,\ell}^\sigma = \omega_\sigma(2n+\ell+3/2) - \mu_\sigma
$
is the eigenvalue and
$
\phi_{n,\ell,m}^\sigma (\mathbf{r}) = R_{n,\ell}^\sigma(r) 
Y_{\ell,m}(\theta_\mathbf{r},\varphi_\mathbf{r})
$
is the eigenfunction. Here, $n$ is the radial quantum number, and $\ell$ and $m$
are the orbital angular momentum and its projection, respectively.
The angular part $Y_{\ell,m}(\theta_\mathbf{r},\varphi_\mathbf{r})$ 
is a spherical harmonic and the radial part is
$
R_{n,\ell}^\sigma (r) = \sqrt{2} (M_\sigma \omega_\sigma)^{3/4} [n!/(n+\ell+1/2)!]^{1/2}
e^{-\bar{r}_\sigma^2/2} \bar{r}_\sigma^\ell L_n^{\ell+1/2}(\bar{r}_\sigma^2),
$
where 
$
\bar{r}_\sigma = \sqrt{M_\sigma \omega_\sigma} r
$ 
is dimensionless and $L_i^j(x)$ is an associated Laguerre 
polynomial. Since $\ell$ and $m$ are good quantum numbers ($\eta \equiv \{\ell, m, \gamma\}$), 
this expansion leads to
$
u_{\ell,m,\gamma} (\mathbf{r}) = \sum_n c_{\ell,\gamma,n} \phi_{n,\ell,m}^\uparrow(\mathbf{r})
$
and
$
v_{\ell,m,\gamma} (\mathbf{r}) = \sum_n d_{\ell,\gamma,n} \phi_{n,\ell,m}^\downarrow(\mathbf{r}).
$

The spherical symmetry of the Hamiltonian simplifies the numerical calculations 
considerably such that the BdG equations reduce to a $2(n_\ell+1) \times 2(n_\ell+1)$ 
matrix eigenvalue problem for a given $\ell$ state
\begin{eqnarray}
\label{eqn:bdg.matrix}
\sum_{n,n'} \left( \begin{array}{cc}
K_{\uparrow,\ell}^{n,n'} & \Delta_\ell^{n,n'} \\
\Delta_\ell^{n',n} & -K_{\downarrow,\ell}^{n',n} 
\end{array} \right)
\left( \begin{array}{c}
c_{\ell,\gamma,n'} \\
d_{\ell,\gamma,n'} 
\end{array} \right)
= \epsilon_{\ell,\gamma} \sum_n 
\left( \begin{array}{c}
c_{\ell,\gamma,n} \\
d_{\ell,\gamma,n} 
\end{array} \right),
\end{eqnarray}
where $n_\ell = (n_c - \ell)/2$ is the maximal radial quantum number 
and $n_c$ is the radial quantum number cutoff to be specified below.
Here, the diagonal matrix element is
$
K_{\sigma,\ell}^{n,n'} = \xi_{n,\ell}^\sigma \delta_{n,n'}
$
and the off-diagonal matrix element is
$
\Delta_\ell^{n,n'} = \int r^2 dr \Delta(r) R_{n,\ell}^\uparrow(r) R_{n',\ell}^\downarrow(r),
$
where $\delta_{i,j}$ is the Kronecker delta.
Furthermore, this procedure reduces the order parameter equation to
\begin{equation}
\label{eqn:op}
\Delta(r) = -g \sum_{\ell,\gamma,n,n'} \frac{2\ell+1}{4\pi} \widetilde{R}_{\ell,\gamma,n}^\uparrow(r) 
\widetilde{R}_{\ell,\gamma,n'}^\downarrow(r) f(\epsilon_{\ell,\gamma}),
\end{equation}
and the local density equations to
\begin{eqnarray}
\label{eqn:nup}
n_\uparrow(r) &=& \sum_{\ell,\gamma,n,n'} \frac{2\ell+1}{4\pi} \widetilde{R}_{\ell,\gamma,n}^\uparrow(r) 
\widetilde{R}_{\ell,\gamma,n'}^\uparrow(r) f(\epsilon_{\ell,\gamma}), \\
\label{eqn:ndo}
n_\downarrow(r) &=& \sum_{\ell,\gamma,n,n'} \frac{2\ell+1}{4\pi} \widetilde{R}_{\ell,\gamma,n}^\downarrow(r) 
\widetilde{R}_{\ell,\gamma,n'}^\downarrow(r) f(-\epsilon_{\ell,\gamma}),
\end{eqnarray}
where we introduced
$
\widetilde{R}_{\ell,\gamma,n}^\uparrow(r) = c_{\ell,\gamma,n} R_{n,\ell}^\uparrow(r)
$
and
$
\widetilde{R}_{\ell,\gamma,n}^\downarrow(r) = d_{\ell,\gamma,n} R_{n,\ell}^\downarrow(r).
$
Notice that the $(2\ell+1)$ factors in Eqs.~(\ref{eqn:op}),~(\ref{eqn:nup}) 
and~(\ref{eqn:ndo}) are due to the degeneracy of each $\ell$ state.
Furthermore, the total number equations become
$
N_\uparrow = \sum_{\ell,\gamma,n} (2\ell+1) c_{\ell,\gamma,n}^2 f(\epsilon_{\ell,\gamma})
$
and
$
N_\downarrow = \sum_{\ell,\gamma,n} (2\ell+1) d_{\ell,\gamma,n}^2 f(-\epsilon_{\ell,\gamma}).
$
These equations generalize the BdG formalism developed in Ref.~\cite{ohashi} to the case 
with unequal masses, unequal chemical potentials and/or unequal trapping potentials.
Having discussed the BdG formalism, next we analyze the ground state ($T = 0$) phases.

First, we analyze the noninteracting ($g = 0$ or $a_F \to 0^-$) case. 
In this case, the discrete energy spectrum can be written as 
$\xi_{n,\ell}^\sigma = \omega_\sigma(n_p + 3/2)$ where $n_p = 2n + \ell$ 
is the principal quantum number. Therefore, for a given $n_p$, 
the orbital angular momentum $\ell$ ranges from $0, 2, ..., n_p$ when 
$n_p$ is even, and it ranges from $1, 3, ..., n_p$ when $n_p$ is odd. 
Since the single pseudo-spin degeneracy 
$
D_{n_p} = \sum_{\ell}^{n_p} (2\ell + 1)
$
of each $n_p$ level is
$
D_{n_p} = (n_p+1)(n_p+2)/2,
$
we can introduce the Fermi level $n_{F,\sigma}$ that corresponds to the 
maximal value of the occupied $n_p$ states at $T = 0$. The condition
$
N_\sigma = \sum_{n_p = 0}^{n_{F,\sigma}} D_{n_p}
$
leads to
$
N_\sigma = (n_{F,\sigma} + 1)(n_{F,\sigma} + 2)(n_{F,\sigma} + 3)/6,
$
and the energy eigenvalue 
$
\epsilon_{F,\sigma} = \omega_\sigma(n_{F,\sigma} + 3/2)
$
that corresponds to the $n_p = n_{F,\sigma}$ state is the Fermi energy 
of the $\sigma$ fermions. For sufficiently large $n_{F,\sigma}$, we notice 
that $\epsilon_{F,\sigma}$ and $N_\sigma$ have a simple relation 
$\epsilon_{F,\sigma} \approx \omega_\sigma(6N_\sigma)^{1/3}$.

At $T = 0$, we can approximately calculate the position $r_*$ where 
the local polarization density $p(r) = |n_\uparrow(r) - n_\downarrow(r)|$ 
becomes zero $p(r_*) = 0$. Using LDA, we find
$
n_\sigma(r) = k_{F,\sigma}^3(r)/(6\pi^2),
$
where
$ 
k_{F,\sigma}(r) = M_\sigma \omega_\sigma (r_{F,\sigma}^2 - r^2)^{1/2}
$
is the local Fermi momentum and 
$
r_{F,\sigma} = (48 N_\sigma)^{1/6}/\sqrt{M_\sigma \omega_\sigma}
$ 
is the Thomas-Fermi radius of the $\sigma$ fermions. This leads to
$
r_* = r_{F,\downarrow} [M_\uparrow \omega_\uparrow/
(M_\uparrow \omega_\uparrow + M_\downarrow \omega_\downarrow)]^{1/2},
$
which is an important length scale because the formation of BCS type Cooper 
pairs is most favored in the momentum space regions when
the Fermi surfaces of $\uparrow$ and $\downarrow$ fermions have minimal 
mismatch, \textit{i.e.} $k_{F,\uparrow}(r_*) = k_{F,\downarrow}(r_*)$. 
Therefore, when $g \to 0^+$, the noninteracting mixture first becomes locally 
unstable against the BCS type superfluidity at $r_*$, as can be seen in our 
numerical calculations which is discussed next.

\begin{figure} [htb]
\centerline{\scalebox{0.55}{\hskip 0mm \includegraphics{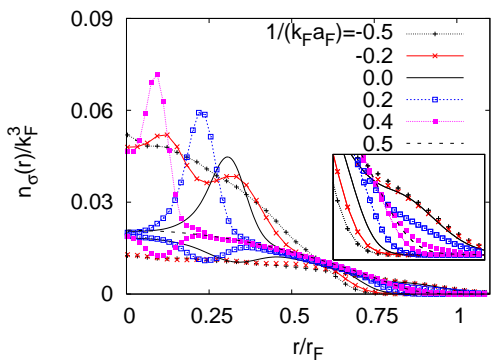}}}
\caption{\label{fig:density}
(Color online) The local density $n_\sigma(r)$ of $^6$Li (with points) and $^{40}$K 
(with lines and points) atoms versus distance $r$ is shown for a population balanced mixture. 
The inset shows the close-up densities near the trap edge.
}
\end{figure}

For this purpose, we solve the BdG equations~\ref{eqn:bdg.matrix},~(\ref{eqn:op}),
~(\ref{eqn:nup}) and~(\ref{eqn:ndo}) self-consistently as a function of the dimensionless 
parameter $1/(k_F a_F)$ where $k_F$ is specified below.
In particular, we consider an equal population mixture of $^6$Li and $^{40}$K 
atoms with $N_\uparrow = N_\downarrow$ and $M_\uparrow = 0.15 M_\downarrow$,
and assume that both $^6$Li and $^{40}$K atoms are trapped with equal trapping 
potentials $V_\uparrow(r) = V_\downarrow(r)$ such that 
$M_\uparrow \omega_\uparrow^2 = M_\downarrow \omega_\downarrow^2$.
We introduce a `reduced' trapping frequency $\omega_r$ via 
$M_r \omega_r^2 = M_\sigma \omega_\sigma^2$, and define an energy scale 
$\epsilon_F$ and two length scales $(k_F, r_F)$ via 
$
\epsilon_F = \omega_r (n_F + 3/2) = k_F^2/(2M_r) = M_r \omega_r^2 r_F^2/2.
$
Notice that $n_F = n_{F,\uparrow} = n_{F,\downarrow}$ when $N_\uparrow = N_\downarrow$. 
Similarly, we define $n_c$ via $\epsilon_c = \omega_r (n_c + 3/2)$. 
In our numerical calculations, we choose $n_F = 15$ and $n_c = 65$,
which correspond to a total of $N = 1632$ fermions and 
$\epsilon_c \approx 4\epsilon_F$, respectively. 
Here, it is important to emphasize that we do not expect any qualitative 
change in our results with higher values of $n_F$ and/or $n_c$,
except for minor quantitative variations.

\begin{figure} [htb]
\centerline{\scalebox{0.55}{\includegraphics{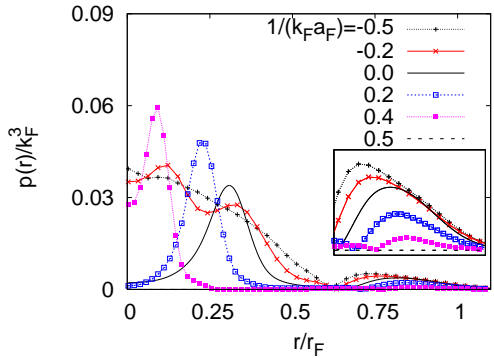} }}
\caption{\label{fig:mag}
(Color online) The local polarization density $p(r) = |n_\uparrow (r) - n_\downarrow (r)|$ 
versus distance $r$ is shown for a population balanced mixture of $^6$Li and 
$^{40}$K atoms. While $^6$Li atoms are in excess near the trap edge, 
$^{40}$K atoms are in excess close to the center.
}
\end{figure}

In Figs.~\ref{fig:op},~\ref{fig:density} and~\ref{fig:mag}, we show the evolution 
of $\Delta(r)$, $n_\sigma(r)$ and $p(r)$ as a function of $1/(k_Fa_F)$. 
For a weakly attracting mixture, while $^{40}$K atoms are in excess around 
the trap center and $^6$Li atoms are in excess close to the trap edge, 
they have similar densities only around 
$r = r_* \approx 0.85 r_{F,\downarrow} \approx 0.61r_F$. 
Therefore, when $1/(k_Fa_F) \lesssim -0.5$, we find that $\Delta(r)$ is spatially 
modulated around $r = 0$, and that its global maximum occurs at $r = r_*$ 
where the mixture is locally unpolarized. The modulation period 
${\cal T} \approx 0.6r_F$ of $\Delta(r)$ is approximately given by
${\cal T} \sim 2\pi / |k_{F,\uparrow} - k_{F,\downarrow}| \approx 0.5r_F$, 
where $k_{F,\sigma} = \sqrt{2M_\sigma \mu_\sigma}$.
As $g$ increases towards unitarity $1/(k_Fa_F) = 0$, we find that the amplitude 
of the modulations dramatically increases around $r = 0$. This is 
because both $n_\uparrow(r)$ and $n_\downarrow(r)$ are highest at $r = 0$, 
which effectively leads to stronger local interactions there since 
$1/[k_F(r)a_F]$ increases with increasing density when $a_F < 0$.
Therefore, the maximum of $\Delta(r)$ eventually occurs at $r = 0$ when the 
effective local interactions become sufficiently strong. 

These spatial modulations of $\Delta(r)$ have dramatic effects on the local 
density of fermions causing pronounced modulations in $n_\sigma(r)$ and $p(r)$ 
close to $r = 0$ as shown in Figs.~\ref{fig:density} and~\ref{fig:mag}.
Since these modulations are significantly large, we hope that they 
can be observed in the future experiments.
Further increasing $g$ towards the molecular limit $1/(k_Fa_F) \gtrsim 0.5$, 
we find that the spatial modulations gradually disappear and the BCS type 
nonmodulated superfluid region expands until the entire mixture becomes 
locally unpolarized. This is expected because the Fermi surfaces disappear 
in this limit, and therefore formation of the molecules does not require matching 
of the Fermi surfaces.

We remark in passing that the BdG equations do not necessarily have a unique 
solution, and depending on the initial values of $\mu_\sigma$ and $\Delta(r)$ 
that are used in the iterative approach, they often yield multiple solutions 
for a given set of parameters. In this manuscript, we show only the physical 
solutions which have lowest energy. Compared to the physical solutions, 
the unphysical ones have considerable qualitative variations around $r = 0$,
\textit{i.e.} both $\Delta(r)$ and $n_\sigma(r)$ have more modulations. 
However, both the physical and the unphysical solutions have very similar
qualitative structure around $r = r_*$, where the BCS type nonmodulated 
superfluidity occurs.

We emphasize that our results are based on the fully quantum mechanical 
BdG formalism, and that they are significantly different from the earlier 
results that are based on the semi-classical LDA method~\cite{pao-mixture, lin-mixture}. 
For instance, these LDA calculations suggest a sharp phase separation in the 
weakly attracting limit and around unitarity such that the BCS type superfluid 
phase is sandwiched between locally polarized normal fermions. 
Therefore, both $\Delta(r)$ and $n_\sigma(r)$ have unphysical discontinuities 
at the normal-BCS superfluid-normal interfaces which indicates breakdown 
of the LDA. This is because the LDA method excludes the possibility of 
a spatially modulated superfluid phase, which is one of the possible 
candidates for the ground state. 
A similar discrepancy between the BdG and the LDA method was previously 
discussed in the context of one-species fermion mixtures with unequal 
populations~\cite{mizushima, torma}. Thus, we conclude quite generally 
that the LDA method is insufficient to describe trapped 
fermion mixtures with mismatched Fermi surfaces, and that it should
be used with caution.

Furthermore, our results for population balanced two-species fermion 
mixtures are qualitatively different from the recent works on one-species 
fermion mixtures with unequal populations~\cite{mizushima, torma}. 
In the latter case, the BCS type superfluid phase occurs around $r = 0$,
and $\Delta(r)$ is spatially modulated towards the trap edge where the 
mixture is locally polarized. Since both $n_\uparrow(r)$ 
and $n_\downarrow(r)$ are very low near the trap edge and the 
modulations have very small amplitudes, it may not be possible to 
observe them at experimentally attainable temperatures. 
However, in our case, spatial modulations with large amplitudes occur 
around $r = 0$ where both $n_\uparrow(r)$ and $n_\downarrow(r)$ are very 
high. These make two-species fermion mixtures very good candidates 
for the observation of spatially modulated superfluid phases in atomic systems.

For instance, spatial modulations of $\Delta(r)$ can be observed by 
using the recently developed technique of spatially resolved radio-frequency 
spectroscopy~\cite{shin-rf}. This technique can be used to locate the nodes
of $\Delta(r)$, since the local quasiparticle excitation spectrum becomes 
gapless at the position of the nodes. In addition, spatial modulations of 
$n_\sigma(r)$ can be observed by using phase-contrast imaging of $^6$Li 
and $^{40}$K populations. In fact, both of these techniques have 
recently been used with great success to characterize the superfluid and 
the normal phases of one-species fermion mixtures with unequal 
populations~\cite{shin-rf}.

In conclusion, we analyzed the ground state phases of harmonically trapped 
$^6$Li-$^{40}$K mixtures with equal populations. In the weakly interacting 
limit and around unitarity, we found that the superfluid order parameter 
is spatially modulated around the trap center. 
Furthermore, we showed that the BCS type superfluidity 
first occurs at a finite distance away from the trap center where the mixture 
is locally unpolarized, and then it gradually expands as the attraction 
strength increases towards the molecular limit until the entire mixture 
becomes locally unpolarized. Since the spatial modulations with large 
amplitudes survive at unitarity, two-species fermion mixtures offer a 
unique opportunity for their observation.

\end{document}